# Dynamic Sounding for Multi-User MIMO in Wireless LANs

Xiaofu Ma, *Student Member,* IEEE, Qinghai Gao, Ji Wang, Vuk Marojevic, and Jeffrey H. Reed, *Fellow*, IEEE

*Abstract*—Consumer electronic (CE) devices increasingly rely on wireless local area networks (WLANs). Next generation WLANs will continue to exploit multiple antenna systems to satisfy the growing need for WLAN system capacity. Multiple-input multiple-output (MIMO) antenna systems improve the spectral efficiency and single user throughput. Multi-user MIMO (MU-MIMO) systems exploit the spatial separation of users for increasing the sum-throughput. In an MU-MIMO system, efficient channel sounding is essential for achieving optimal performance. The system analysis in this paper provides insights into the rate at which to perform channel sounding. This paper shows that optimal sounding intervals exist for single user transmit beamforming (SU-TxBF) and MU-MIMO, and proposes a low-complexity dynamic sounding approach for practical MU-MIMO WLAN deployments. The proposed approach adjusts the sounding interval adaptively based on the real-time learning outcomes in the given radio environment. Using real over-the-air channel measurements, significant throughput improvements (up to 31.8%) are demonstrated by adopting the proposed dynamic sounding approach, which is compliant with IEEE 802.11ac[1].

*Index Terms* — **Multi-user MIMO, wireless local area network, dynamic sounding, channel state information.**

## I. Introduction

Wireless local area networks (WLANs) have experienced dramatic growths during the last decade with the proliferation of IEEE 802.11 devices [1]. WLAN continues to be the dominating wireless infrastructure for wireless consumer electronic (CE) devices with various applications [2], [3], [4]. IEEE 802.11 WLAN Task Groups are pursing gigabit wireless communications to further increase the throughput and

Manuscript received March 29, 2017; accepted June 16, 2017. Date of publication July 20, 2017.[1] This work was supported in part by Wireless@ Virginia Tech. *(Corresponding author: Xiaofu Ma.)*

Xiaofu Ma is with the Virginia Polytechnic Institute and State University, Blacksburg, VA 24060 USA (e-mail: xfma@vt.edu).

Qinghai Gao is with Qualcomm Atheros, San Jose, CA 95110 USA (e-mail: qinghai.gao@gmail.com).

Ji Wang is with the Virginia Polytechnic Institute and State University, Blacksburg, VA 24060 USA (e-mail: traceyw@vt.edu).

Vuk Marojevic is with the Virginia Polytechnic Institute and State University, Blacksburg, VA 24060 USA (e-mail: maroje@vt.edu).

Jeffrey H. Reed is with the Virginia Polytechnic Institute and State University, Blacksburg, VA 24060 USA (e-mail: reedjh@vt.edu).

spectral efficiency for the growing number of CE devices [5].

A typical multiple antenna WLAN system includes one access point (AP) and several mobile stations. APs have been MIMO capable since the release of IEEE 802.11n [6]. The new IEEE 802.11 standards, IEEE 802.11ac [7] and 802.11ax [8], add multi-user MIMO (MU-MIMO) as the key technology to improve system throughput. In particular, an MU-MIMO system enables a MIMO capable AP to communicate with multiple users simultaneously through spatial diversity exploitation and spatial multiplexing. Compared with the traditional MIMO system, the use of MU-MIMO systems reduces the requirements of the end user devices, which do not need to have multiple antennas.

However, the deployment of MU-MIMO systems faces several challenges. This paper addresses one such challenge related to acquiring accurate channel state information (CSI) that describes the properties of the communication link. Inaccurate CSI can mislead the beam steering at the transmitter and therefore results in severe cross-user interference at the receiver, causing significant throughput degradation. Explicit channel sounding is necessary to acquire CSI in an MU-MIMO capable WLAN system. During one explicit channel sounding process, a pre-known sequence of data is transmitted to the receivers on the target channel. Based on the received signal, the receiver estimates the CSI and sends this information back to the transmitter. The CSI accuracy depends not only on the correctness of the estimation process, but also on the sounding update frequency. However, the increase of channel sounding frequency reduces the data transmission windows. This creates a trade-off between collecting up-to-date CSI and maximizing the transmission time.

Deriving a suitable sounding interval for an MU-MIMO system in a real scenario is challenging because of the dynamic and unpredictable channel conditions. For example, in the channel environments with static devices and few human activities, the channel stays relatively stable, which requires less frequent channel sounding. In contrast, device and environment mobility can create a fast changing channel environment, which would dramatically degrade the system throughput unless the CSI is frequently updated. Therefore, the sounding-transmission trade-off needs to be continuously evaluated.

This paper extends the state of the art in channel sounding for MU-MIMO systems by providing a theoretical analysis and experimental results based on extensive over-the-air channel measurements in an indoor WLAN environment. The results show how dynamic sounding can provide better throughput over static sounding at minimal computational and signaling overhead. The main contributions are summarized as follows.

(1) The selection of the channel sounding frequency is characterized and evaluated for single user transmit beamforming (SU-TxBF) and MU-MIMO in WLAN. The mathematical formation is used to formally describe the problem and explain the performance trend. Rather than solving the problem analytically, this paper presents a practical approach: Extensive channel measurements were conducted in both, static and dynamic indoor scenarios using MU-MIMO test nodes to demonstrate the existence of a channel-specific optimal sounding interval.

(2) A dynamic sounding approach of low complexity is derived to improve the effective throughput in real WLAN indoor environments. Using a developed 802.11ac emulator seeded with measured channel information, the proposed approach is shown to achieve throughput improvements of up to 31.8%.

The rest of the paper is organized as follows: Section II presents the problem and related work. In Section III, the trade-off between sounding overhead and throughput is evaluated. A practical dynamic sounding approach is proposed and evaluated in Section IV. Section V concludes the paper.

## II. PROBLEM STATEMENT AND RELATED WORK

### A. Problem Statement

MU-MIMO is a promising technology for increasing the system throughput of WLAN systems. A suitable channel sounding strategy is critical for this. Too infrequent sounding operations cause more cross-user interference because of the out-dated CSI, whereas too frequent sounding operations unnecessarily take up the effective transmission time, both leading to performance deficits. The problem thus consists in finding a suitable operational point that trades the CSI accuracy for effective transmission time as a function of the radio environment. In the context of IEEE 802.11ac and future WLAN systems, this paper assesses the following two aspects related to explicit channel sounding:

(1) What is the optimal sounding interval that maximizes system throughput?

(2) How to dynamically adapt the sounding interval for rapidly changing radio environments?

### B. Background and Related Work

To increase the MU-MIMO system throughput, many research efforts focus on the strategy of user grouping. These works assume that perfect CSI information is available at the transmitter. The scheme presented by Dimic and Sidiropoulos [9] greedily groups users based on the estimated capacity for improved system throughput. Channel orthogonality can also be an effective metric to group users. For example, a heuristic approach was proposed by Yoo and Goldsmith [10] to gather the most orthogonal users in one group and disperse highly-correlated users over different time slots. The use of group membership and group identifiers for managing MU-MIMO downlink transmissions [11] was studied specifically for WLAN systems.

The media access control (MAC) protocol design is another research area for MU-MIMO systems. A modified CSMA/CA protocol [12] for MU-MIMO systems was proposed, focusing on ACK-replying mechanisms to improve system throughput, where CSI is assumed to be known for the minimum mean square error (MMSE) precoding. The proportional fair allocation mechanism for MU-MIMO transmission was adopted by Valls and Leith [13] and assumes that the AP has full knowledge of the channel. These works have paved the path for practical implementations of MU-MIMO in WLAN systems; however, they do not evaluate the channel sounding overhead and CSI imperfection, which directly affect MU-MIMO throughput.

Some perspectives of CSI have been analysed for MU-MIMO in commercial wireless systems using numerical evaluation. Training and scheduling aspects of MU-MIMO schemes [14] that rely on the use of outdated CSI were investigated, showing that certain MU-MIMO throughput can be achieved even with outdated CSI. For the downlink of MU-MIMO-based FDD systems, the channel feedback mechanism [15] was studied to consider user diversity and the channel correlations in both time and frequency.

Recent literature also provides experimental results for MU-MIMO. Balan et al. [16] demonstrated that the system data rate of the MU-MIMO system using channel feedback was much better than without channel feedback. The impact of CSI compression on the feedback overhead was investigated by Xie et al. [17]. In order to reduce the sounding overhead, implicit sounding [18] may outperform explicit sounding with lower time overhead. However, implicit sounding requires more computation for both the channel and transceiver radio frequency (RF) chain calibration to maintain full channel reciprocity. In addition, imperfect CSI at the transmitter degrades the throughput of a MU-MIMO system throughput more severely than that of a basic MIMO system. The standard form of channel sounding in IEEE 802.11ac is only explicit [19], which requires the use of channel measurement frames.

The non-negligible overhead produced by explicit channel sounding motivates research on finding the optimal sounding frequency. The relationship between sounding frequency and MU-MIMO system throughput needs both theoretical and experimental investigations. There are theoretical contributions for deriving the optimal sounding interval for a basic MIMO system. Zhang et al. [20], for instance, assumed the Rayleigh block-fading channel. These solutions cannot be directly applied to the MU-MIMO system. In addition, in practical deployments where environmental or device mobility can cause significant channel variations, a predefined sounding interval would not be suitable. The current literature lacks system-level analyses and evaluations that consider dynamic sounding frequency as another degree of freedom for practical WLAN systems. This paper analyzes and experimentally validates the importance of dynamic channel sounding for emerging and future WLAN deployments.

## III. STATIC OPTIMAL SOUNDING INTERVAL

A well-chosen sounding interval maximizes the effective spectral efficiency, which is defined as the total data rate delivered to all stations over the transmission bandwidth. This effective spectral efficiency depends on two major factors: the effective transmission percentage and the instantaneous data rate as a function of time. This section uses the effective spectral efficiency to analyze the performance of SU-TxBF and MU-MIMO, both being part of the new generation WLAN systems.

The goals of this section are (1) to show the existence of a channel-specific sounding interval using the over-the-air measurements, and (2) to demonstrate the dependency of this static optimal sounding interval on the different system parameters. The static optimal sounding interval is the one among all possible constant sounding intervals that leads to the highest effective spectral efficiency.

The channel model is presented first, followed by the optimization problem formulation and evaluation using an IEEE 802.11ac emulator seeded with real channel measurements collected for different indoor scenarios.

### A. Channel Model

In this paper, the impulse response of the channel at time $t$ is denoted by $h(t, \tau)$, where $\tau$ is the delay variable. Due to the multipath fading effect, $h(t, \tau)$ can be expressed as

$$h(t, \tau) = \sum_{k=1}^{N} a_k(t) \delta(\tau - \tau_k) \qquad (1)$$

where $N$ represents the number of taps, $a_k$ represents the weight of the tap $k$ and $\tau_k$ represents the delay on the $k$ th tap. The impulse responses are based on the cluster model [18], and the power of delayed responses decays linearly on a log-scale, i.e., the power decays exponentially. The powers of the taps in overlapping clusters are summed at each delay. The frequency domain representation is adopted for MIMO operation analysis. The system analysis is presented for only one single subcarrier for notation simplicity, but the analysis applies to any number of subcarriers.

### B. Effective Spectral Efficiency

The time consumed for each data transmission period is denoted by $T_\Delta$ and includes two parts: channel sounding and data transmission. $T_\Delta$ can also be viewed as the time gap between two successive sounding operations. The operation time for a sounding operation is denoted by $T_S$. The value of $T_S$ differs for different number of user(s) served by the SU-TxBF or MU-MIMO transmission. The effective transmission time percentage then becomes $(1 - T_S / T_\Delta)$. It is assumed that the channel undergoes block fading and each block lasts for a very small time $\Delta_t$. Thus, the effective spectral efficiency $C_e(T_\Delta)$ during $T_\Delta$ can be formulated as

$$C_e(T_\Delta) = (1 - \frac{T_S}{T_\Delta}) \cdot \frac{1}{M} \cdot \sum_{m=1}^{M} C(T_S + m \cdot \Delta_t). \qquad (2)$$

Here, $M = \lfloor T_\Delta / \Delta_t \rfloor$, and $C(t)$ is the spectral efficiency. $C(t)$ is measured as the expected value over channel realization and transmission of $\sum_{i=1}^{N_{ss}} log_2(1 + \gamma_i(t))$, where $N_{ss}$ is the number of transmission streams and $\gamma_i(T_\Delta)$ represents the SNR of the $i$ th spatial stream. It is obvious from (2) that $(1 - T_S / T_\Delta)$, the first part of $C_e(T_\Delta)$, increases with increasing $T_\Delta$. The value of the second part of the equation, $\frac{1}{M} \cdot \sum_{m=1}^{M} C(T_S + m \cdot \Delta_t)$, also changes with $T_\Delta$. The problem formulation for the optimal sounding interval can therefore be written as

$$max \ C_e(T_\Delta) \ s.t. \ T_\Delta \geq T_S. \qquad (3)$$

In order to maximize the effective spectral efficiency with respect to the channel sounding interval $T_\Delta$, it is necessary to analyze the impact of $T_\Delta$ on $\gamma_i(T_\Delta)$.

The number of transmit antennas is denoted by $N_t$. For SU-TxBF using zero-forcing precoding, the SINR of the spatial stream $i$ after $t_\Delta$ from the channel estimation time $t_0$ can be expressed as (without losing generality, $t_0$ is set to 0)

$$\gamma_i(t_\Delta) = \frac{P_i |h_{s(i)}^*(t_\Delta) \cdot w_{s(i)}(t_0)|^2}{N_0 + \sum_{j \neq i} P_j |h_{s(i)}^*(t_\Delta) \cdot w_{s(i)}(t_0)|^2}, \qquad (4)$$

where $P_i$ is the power allocated to spatial stream $i$, $h_{s(i)}(t_\Delta)$ is the 1-by- $N_t$ frequency-domain channel response vector of spatial stream $i$ at time $t_\Delta$, $w_{s(i)}(t_0)$ is the $N_t$-by-1 streering vector for spatial stream $i$ based on the channel estimation at time $t_0$, and $N_0$ is the channel noise power. The expression $\sum_{j \neq i} P_j |h_{s(i)}^*(t_\Delta) \cdot w_{s(i)}(t_0)|^2$ is the term of cross-stream interference for stream $i$ at the receiver.

In the MU-MIMO case, when zero-forcing precoding is used for multi-stream steering, the SINR of the spatial stream $i$ for user $n$ after $t_\Delta$ from the channel estimation time $t_0$ is

$$\gamma_{n,i}(t_\Delta) =$$

$$\frac{P_{n,i}\left|\boldsymbol{h}_{n,i}^*(t_\Delta)\cdot\boldsymbol{w}_{n,i}(t_0)\right|^2}{N_0 + \sum_{j\neq i}P_{n,j}\left|\boldsymbol{h}_{n,i}^*(t_\Delta)\cdot\boldsymbol{w}_{n,j}(t_0)\right|^2 + \sum_{m\neq n}P_{n,j}\left|\boldsymbol{h}_{n,i}^*(t_\Delta)\cdot\boldsymbol{w}_{m,j}(t_0)\right|^2},$$

$$(5)$$

where $P_{n,i}$ is the power allocated to stream $i$ for user $n$, $\boldsymbol{h}_{n,i}(t_\Delta)$ is the 1-by-$N_t$ frequency-domain channel response vector of spatial stream $i$ for user $n$ at time $t_\Delta$, and $\boldsymbol{w}_{n,i}(t_0)$ is the $N_t$-by-1 streering vector for spatial stream $i$ at user $n$ based on the channel estimation at time $t_0$. The expression $\sum_{j\neq i}P_{n,j}\left|\boldsymbol{h}_{n,i}^*(t_\Delta)\cdot\boldsymbol{w}_{n,j}(t_0)\right|^2$ in (5) is the cross-stream interference at user $n$, and $\sum_{m\neq n}P_{n,j}\left|\boldsymbol{h}_{n,i}^*(t_\Delta)\cdot\boldsymbol{w}_{m,j}(t_0)\right|^2$ is the cross-user interference.

The optimal solution of the formulation in (3) would maximize the effective spectrum efficiency. However, this problem is very complex to solve for a given channel and is impractical as the channel status continuously changes in a real environments. It is therefore infeasible to derive the closed-form solution for IEEE 802.11ac consumer electronic systems. Instead, the next subsection evaluates the transmission trade-off using channel measurements in different indoor environments as a practical approach to the problem.

### C. Evaluation of Trade-off Between Accurate CSI and Sounding Overhead

#### (1) Experimental System Setup

In this section, the dependency of the optimal sounding interval on the different system parameters is evaluated in a WLAN-based MU-MIMO environment. Consider a typical WLAN topology with one AP and 12 stations. The AP is equipped with four antennas and each station is equipped with a single antenna. The system can serve up to three stations simultaneously on the downlink, which is typical for a practical WLAN scenario.

The measurements of the actual channel information were conducted using four test nodes in an office environment as illustrated in Fig. 1. Each test node was equipped with four antennas and a 4×4 four-stream 802.11ac transceiver chipset. The over-the-air transmissions were conducted over a 40 MHz channel. One test node operated as the transmitter AP, and the remaining three operated as the receiver stations. The 12 receive antennas were spaced apart from one another and distributed randomly in the office. Those 12 sets of channel samples are used to mimic the channel from one four-antenna AP to 12 single-antenna stations.

The channel measurements were collected under two scenarios, the *low Doppler* and *high Doppler* scenarios. The low Doppler scenario refers to a low-mobility radio environment with no human activity or device movement during the channel measurements. Channel samples under the high Doppler scenario were collected when significant human movement existed. Since the antenna movement and human movement are relative, these channel samples were used to evaluate the high dynamic environment. Notice that there are still random variations in both the low and the high Doppler scenarios.

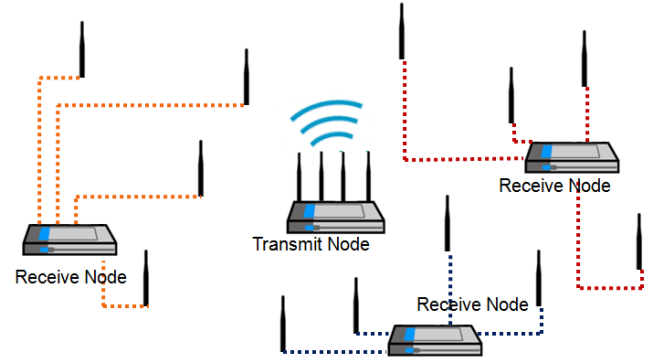

Fig. 1 Measurement setup sketch map

With the measured channel information, trials with different sounding intervals can be conducted using identical channels for comparison. A measurement-driven MU-MIMO-OFDM emulator was rigorously implemented according to the IEEE 802.11ac specifications. This emulator was seeded with the over-the-air channel information to test the performance of the MAC and physical (PHY) layer algorithms. Singular value decomposition (SVD) was employed as the precoding scheme, and the MMSE receiver was implemented to mitigate the cross-user interference.

Since the focus of this paper is the selection of the sounding interval, other possible effects on the system performance have been carefully eliminated. In particular, it is assumed that the AP adapts the optimal modulation and coding scheme (MCS) during each simulation trial. The evaluation parameters of the IEEE 802.11ac system are summarized in Table I.

TABLE I
EVALUATION PARAMETERS

| Parameters | Values |
|---|---|
| Maximum AMPDU Duration | 2 ms |
| MPDU Length | 1556 bits |
| MSDU Length | 1508 bits |
| SIFS Duration | 0.016 ms |
| Bandwidth | 40 MHz |
| Number of OFDM Subcarriers for Data | 108 |
| Guard Interval for OFDM Symbol | 400 ns |

In a single user transmission scenario, the sounding overhead of 802.11ac comes from the explicit feedback mechanism, i.e., the information exchange between the AP and the mobile stations. This information exchange during each channel sounding process is as follows: The AP first broadcasts a Null Data Packet Announcement (NDPA). After a Short Interframe Space (SIFS) time interval, the AP then sends out a Null Data Packet (NDP) to sound the transmission channel. From the station's perspective, after receiving the NDP from the AP, it waits for a SIFS interval and responds to the AP with a compressed beamforming (CBF) report. Hence, the time required for each channel sounding operation for single user transmit beamforming can be expressed as

$$T_{S\_SU} = T_{NDPA} + T_{SIFS} + T_{NDP} + T_{SIFS} + T_{CBF} + T_{SIFS}, \quad (6)$$

where $T_{NDPA}$, $T_{SIFS}$, $T_{NDP}$ and $T_{CBF}$ are the time durations for the NDPA transmission, SIFS, NDP transmission and CBF transmission, respectively. Notice that NDPA and NDP are used by the AP to control the sounding interval which does not need to be known by the stations.

The sounding duration for MU-MIMO in WLAN differs from that of the single user case. It also includes the beamforming report poll(s) for all stations as well as a sequence of the stations' CBF reports. The sounding procedure for MU-MIMO initiates exactly the same as the SU-TxBF sounding procedure. The AP first broadcasts an NDPA. After a SIFS time interval, the AP sends out an NDP to sound the transmission channel. However, to retrieve the channel information from each STA, the MU sounding procedure needs one or more Beamforming Report Poll (BRP) frames to collect responses from all the stations, one by one. A poll frame is not needed for the first station, which is specified in the NDPA frame, but starting from the second and subsequent stations, BRP frame for each station is required. The AP will integrate all the received CBF reports together into a steering matrix. Therefore, the time required for one channel sounding operation under a MU-MIMO transmit beamforming can be derived from (6) and expressed as

$$\begin{aligned} T_{S\_MU} = {}& T_{NDPA} + T_{SIFS} + T_{NDP} + T_{SIFS} \\ & + N_u \cdot (T_{SIFS} + T_{CBF}) + (N_u - 1) \cdot (T_{SIFS} + T_{BRP}) \end{aligned}, (7)$$

where $N_u$ represents the number of users in the multi-user transmission group, and $T_{BRP}$ is the time duration for a BRP frame transmission.

### (2) Evaluation Results

Fig. 2 plots the output SINR as a function of the sounding interval for SU-TXBF transmission, MU-MIMO transmission with 2 users (MU2), and MU-MIMO transmission with 3 users (MU3). More precisely, the SINR for SU, MU2, and MU3 cases are compared with sounding intervals ranging from 2 ms to 400 ms in low (Fig. 2a) and high (Fig. 2b) Doppler scenarios. The SINR for SU-TxBF remains nearly constant. As opposed to MU-MIMO, SU-TxBF does not need to deal with inter-user interference. As a result, the channel estimation is demonstrated here to be fresh enough even for a 400 ms sounding gap for SU-TxBF. The SINR would eventually degrade when the gap increases further.

MU-MIMO is more sensitive to inaccurate CSI because the spatial stream steering is done in such a way that the inter-user interference is minimized for MU transmission as opposed to maximizing the user's SINR for SU-TxBF. Fig. 2 correspondingly shows how the SINR value for MU-MIMO transmission quickly decreases with the sounding gap even in the low Doppler scenario (Fig. 2a) and more quickly for the high Doppler scenario (Fig. 2b). This is because the CSI accuracy for beam steering degrades with increasing time gap between the channel estimation and the data transmission. It is also observed that the SINR per spatial stream decreases with the increasing number of users in the MU transmission in the system. For example, after 200 ms since the last channel sounding operation, the SINR has dropped by 4 dB (from 31 to 27 dB) for MU2 and by 5 dB (from 27 to 22 dB) for MU3 in the low Doppler scenario. The SINR degradation is much more severe for the high Doppler scenario: At 200 ms, for MU2 and MU3, the SINR values are 14 and 8 dB. These results illustrates the significant impact of fast changing channels on SINR for MU transmission and the relative robustness for SU-TxBF.

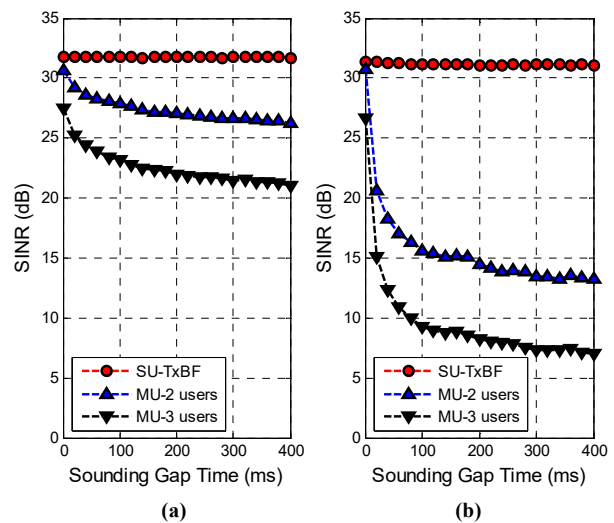

Fig. 2. SINR as a function of sounding interval in (a) the low Doppler scenario and (b) the high Doppler scenario.

Fig. 3 shows the corresponding total PHY layer throughput as a function of the sounding gap. For both the low (Fig. 3a) and high (Fig. 3b) Doppler scenario, the total PHY layer throughput stays relatively constant for SU-TxFB. This is because the sufficiently high SINR for SU-TxFB can support the highest MCS, which leads to constant throughput. This confirms to the SINR results shown in Fig. 2a and Fig. 2b.

On the other hand, the throughput decreases monotonically for MU2 and MU3 transmissions. This illustrates the impact of cross-user interference due to the outdated CSI on system performance. The drop of the system throughput under the same sounding gap is enlarged by the number of users within MU transmission. Specifically for the low Doppler scenario, when the sounding gap increases from 2 to 50 ms, the PHY throughput of MU3 drops by 18% compared to only 5.6% drop for MU2. This is because the increased number of users within an MU transmission worsens the cross-user interference. The same phenomenon can be also observed from the high Doppler scenario.

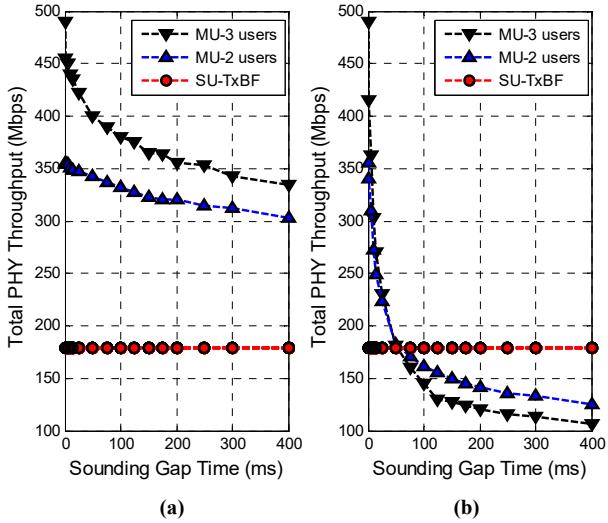

**(a)**                    **(b)**

Fig. 3. Total PHY throughput as a function of sounding interval in (a) the low Doppler scenario and (b) the high Doppler scenario.

Compared with the low Doppler scenario, the throughput drop in the high Doppler scenario as the sounding interval increases becomes more significant as shown in Fig. 3b. When the sounding gap is less than 10 ms, the PHY layer throughput of MU3 outperforms that of MU2, and both outperform SU-TxBF. When the sounding gap approaches 50 ms, the PHY layer throughputs of MU2 and MU3 drop to the same level as SU-TxBF. Beyond 50 ms, MU2 has a higher total PHY layer throughput than MU3 and both are below SU-TxBF throughput.

It is also observed from Fig. 3 that for the same length of sounding gap, a throughput difference exists between the low and high Doppler scenarios, and this difference is more significant when the number of users within MU transmission is larger. For example, when the sounding gap is 50 ms, the throughput difference for MU2 between high and low Doppler scenarios is 165 Mbps, whereas the difference becomes 220 Mbps for MU3. When the sounding gap increases to 100 ms, the difference becomes 180 Mbps and 245 Mbps, respectively.

When the sounding overhead is considered, the throughput is the actual throughput that can be achieved and delivered to the end users. This is called effective throughput or MAC throughput. Fig. 4 shows the effective throughput as a function of the sounding gap for MU3. All curves illustrate a similar trend: when the sounding gap is small, the sounding overhead dominates the air time, which causes the low effective throughput. When the sounding gap increases, the relative sounding overhead correspondingly decreases and the throughput increases until a point when the outdated CSI dominates. Note that the PHY layer throughput, which does not account for the sounding overhead, decreases over time since the last sounding operation because of the decreasing accuracy of the available CSI over time.

The optimal sounding interval is around 50 ms for the low Doppler scenario (Fig. 4a). Since the channel environment is relatively stable, the outdated channel information does not significant affect the throughput even when the sounding interval becomes as large as 100 ms or more. This is different for the high Doppler scenario (Fig. 4b), where the MAC throughput quickly drops when passing the optimal sounding interval because of the inter-user interference caused by the rapidly changing environment and inaccurate CSI. The highest throughput for the high Doppler case is achieved for a 10-20 ms sounding interval.

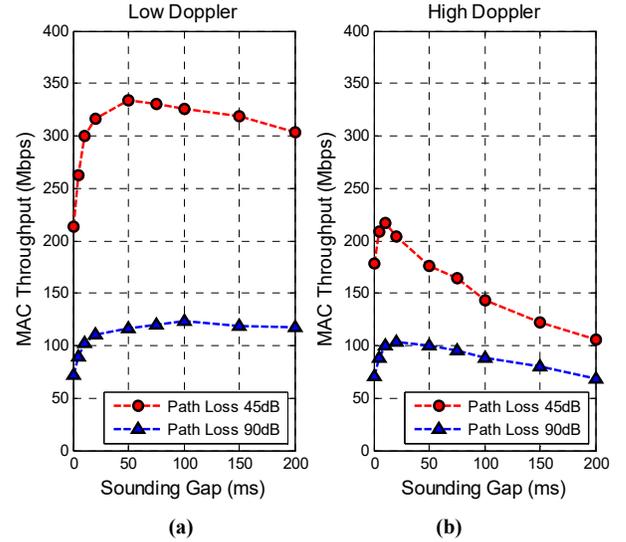

**(a)**                    **(b)**

Fig. 4. Effective throughput (MAC throughput) as a function of the sounding interval under different path losses for (a) the low Doppler scenario and (b) the high Doppler scenario.

It is observed from Fig. 4 that the optimal sounding interval varies not only under different channel variation conditions, but also under different instantaneous path losses. In particular, the effective throughout degrades as the pass loss increases in either low or high Doppler scenario. This is reasonable as larger pass loss leads to lower SINR. In the low Doppler scenario, the optimal sounding interval is 50 ms if the path loss is 45 dB, whereas the optimal value becomes around 100 ms if the path loss is 90 dB. A similar phenomenon can be observed in the high Doppler scenario, where the optimal sounding interval under 45 dB pass loss is around 10 ms, which is lower than that under 90 dB pass loss (around 20ms).

Through the over-the-air measurements, this section shows the existence of the channel-specific optimal sounding interval and its dependency on the channel variation and path losses. The channel variation indicates the device or environment motion, whereas the path loss reflects the geolocation of both the AP and the station.

The SINR and Doppler profile of each station's spatial stream may change rapidly based on the instantaneous channel environment. Since the motion in the environment and the geolocation of the stations are dynamic and unpredictable, it is necessary to design an algorithm that dynamically determines a suitable operational point that trades CSI accuracy for effective transmission time in real time as a function of the radio environment.

## IV. Sounding in Dynamic Environments

So far, the determination of the optimal sounding interval has been investigated in a radio environment characterized by a low or high Doppler scenario. In dynamic indoor environments, the channel conditions change unpredictably because of the random movement of human and wireless devices, or other changes to the environment. The sounding interval requires dynamic adaptation based on the instantaneous channel conditions. Thus, a dynamic sounding approach is designed that (1) is aware of the instantaneous channel condition, (2) infers the suitable sounding interval, and (3) adapts the sounding interval. This is enabled by continuously estimating the statistics of the instantaneous channel conditions, calculating a reference throughput, and providing internal feedback when the reference throughput drops. The reference throughput indicates the accumulated MAC layer throughput since the last sounding operation.

### A. Proposed Dynamic Sounding Approach

To trace the time variation of the radio environment, a Doppler spectrum profile can be maintained from the feedback CSI to estimate the optimal sounding interval. However, both the Doppler spectrum profile creation and the theoretical calculation of the optimal sounding interval require a lot of computing power and time, which is very challenging for the real-time system implementations. To simplify the system design, a sounding adjustment is designed such that it depends on the accumulated data rate, which can be easily collected at the AP.

Both the sounding overhead and the instantaneous transmission conditions are considered to estimate the time correlation of the channel variation and to calculate the effective throughput. The accumulated effective throughput is used as the reference throughput to adjust the sounding interval. All 802.11ac data frames are sent in an aggregated MAC protocol data unit (AMPDU). The reference throughput is calculated after the $n$ th AMPDU transmission since the last sounding operation using

$$R_{TH}(n) = \frac{\sum_{m=1}^{n} \sum_{j=1}^{N_u} D(m,j)}{T_s + \sum_{m=1}^{n} T_{AMPDU}(m)}, \qquad (8)$$

which represents the ratio between the estimate of the accumulated successfully-transmitted AMPDUs during the downlink MU transmission and the time period. This period includes the overhead of the last sounding operation and the time for accumulated downlink AMPDU transmissions. $D(m,j)$ stands for the successfully transmitted data amount for the $m$ th AMPDU of user $j$ since the last sounding operation, and $T_{AMPDU}(m)$ is the time duration of the $m$ th downlink AMPDU transmission.

The basic idea of the dynamic sounding approach is to trigger the sounding operation when the reference throughput starts degrading. In the proposed dynamic sounding approach, which is formally described in Algorithm 1, the AP regularly estimates the reference throughput as a function of time since the last sounding operation. After a sounding operation, the PHY layer throughput is expected to improve and then degrade over time until the next sounding operation. The reference throughput, on the other hand, initially increases as the sounding overhead dominates the effective throughput. However, the reference throughput starts dropping as the increasing CSI inaccuracy over time becomes the dominant factor degrading the effective throughput, which continues degrading until there is another sounding operation.

Since the reference throughput does not reflect the instantaneous channel statistic, but, rather, an accumulated value representing a time interval, the impact of the instantaneous fluctuation of noise on triggering a sounding operation is mitigated. The proposed approach also works when uplink transmission exists, since the reference throughput is a metric that reflects the throughput trend of the downlink transmission over time. Algorithm 1 provides the details of the proposed dynamic sounding approach. From (8), it can be seen that the algorithm complexity is low and, hence, well suited for commercial MU-MIMO systems.

---

**Algorithm 1.** Dynamic Sounding Approach

---

***Stage 1:  Initialization***

1    Flag for whether sounding is needed:  $s = 1$ .

2    Flag for whether the current AMPDU packet is the first one after sounding:  $f = 0$ .

3    The transmitted AMPDU index:  $n = 0$ .

***Stage 2:  MU Downlink Operation***

4    **while** MU Downlink Transmission

5        **if**   $s = 1$

6            Operate channel sounding.

7            $f = 1$ .

8        **end if**

9        Transmit AMPDU using rate adaptation algorithm.

10        $n = n + 1$ .

11        Update  $R_{TH}(n)$  using (8)

12        **if**   $f = 1$  or  $R_{TH}(n-1) < R_{TH}(n)$

13            Set $s = 0$ and prepare for sending the next AMPDU.

14            $f = 0$ .

15        **else**

16            Set  $s = 1$ and prepare for sounding.

17        **end if**

18    **end while**

---

### B. Evaluation

Consider a WLAN system with one AP and 12 stations. The 802.11ac emulator discussed in Section III with the over-the-air channel measurements is used as the evaluation platform. Since the system can support up to three stations for MU-MIMO transmission, it is considered that any three stations out of the 12 are grouped.

The performance of the proposed approach is evaluated using the collected channel measurements, including the low and high Doppler scenarios. The dynamic scenario is mimicked by alternating between the low and high Doppler channel samples. Specifically, the channel samples are rearranged into an alternation between high and low Doppler conditions with alternating 50ms of high and low Doppler channel samples.

As discussed in Section II, contributions exist in literature that derive the optimal static sounding interval for a basic MIMO system assuming the Rayleigh block-fading channel, such as [20]. These solutions are derived for a basic MIMO system, and cannot be directly applied to MU-MIMO systems. In addition, in a practical scenario where environmental or device mobility can cause significant and unpredictable channel variations, where the pre-computed sounding interval may not be efficient.

In order to experimentally validate the importance of dynamic channel sounding for emerging and future WLAN deployments, we compare the proposed approach with two static schemes that correspond to the state-of-the-art solutions presented in [20]. These two schemes—one has a short and the other a long sounding interval—provide the optimal fixed sounding intervals for the low and high Doppler scenarios. We call them low Doppler approach (LDA) and high Doppler approach (HDA). Notice that the channel variation level cannot be predicted in a real radio environment, so it is not feasible to know the optimal fixed sounding interval. Fig. 5 illustrates the sounding events for the three approaches and Fig. 6 compares the performance.

Fig. 5 indicates the time stamps for channel sounding as triggered by the AP using the three schemes in an example 400 ms period with alternating high-low Doppler conditions. As expected, under both conditions, the sounding interval under HDA is always around 11ms and the sounding interval under LDA is around 43 ms. In contrast, the sounding interval of the dynamic sounding approach varies according to the change of the radio environment, which illustrates its channel awareness functionality. Particularly, the average sounding interval of the proposed approach in the high Doppler scenario is around 11 ms, which increases to around 41 ms for the low Doppler case. As can be observed from Fig. 5, the operation activity of the dynamic sounding approach stays almost the same as that of LDA in low Doppler scenario and that of HDA in the high Doppler scenario. This illustrates the accuracy of the channel awareness process of the proposed dynamic sounding approach.

The throughput improvement of approach $X$ over approach $Y$ is defined as $(R_X - R_Y)/R_Y$, where $R_X$ and $R_Y$ are the throughputs achieved by approach $X$ and $Y$. Fig. 6 shows the throughput improvement of the proposed dynamic sounding approach over LDA (green bars) and over HDA (yellow bars) in the high Doppler, low Doppler, and alternating conditions.

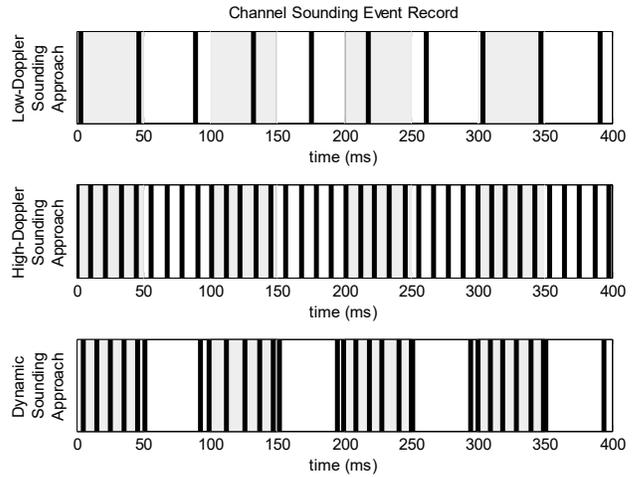

Fig. 5. Example sounding event records for the two optimal static sounding approaches, LDA (top) and HDA (middle), and the proposed dynamic sounding approach (bottom). Grey shaded areas indicate the high Doppler scenario and white areas the low Doppler scenario. The black solid lines illustrate the sounding events.

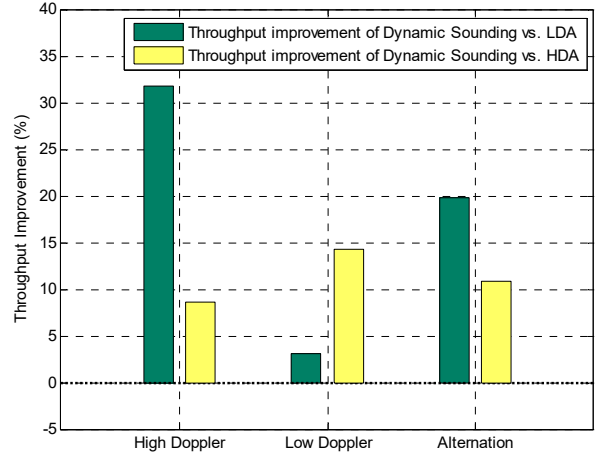

Fig. 6. Throughput improvement of dynamic sounding over LDA and HDA.

It is observed that in the high Doppler scenario, the throughput improvement of dynamic sounding over LDA is as high as 31.8%. This illustrates the benefit of operating frequent sounding in rapidly changing radio environments. In addition, the throughput improvement of the proposed dynamic sounding approach over HDA is 8.6%, which means that the dynamic sounding approach outperforms the scheme that uses the optimal fixed-sounding interval. This is because in the high Doppler scenario the channel variation levels change over time, and dynamic sounding approach can capture the channel changes timely to trigger a sounding operations.

Similarly, it is observed that there is a 14.3% improvement of using dynamic sounding approach over HDA in the low Doppler scenario. Dynamic sounding and LDA achieve similar performance, where the difference is 3.1%. In the alternation conditions, however, none of the fixed approaches achieve the same level of performance as the proposed solution. In particular, dynamic sounding outperforms static sounding by 19.8% (LDA) and 10.9% (HDA). The proposed approach explores the change in the channel environment and trades the CSI accuracy for effective transmission time as the function of observed channel variations.

## V. Conclusions

This paper has presented a framework for evaluating channel sounding intervals with practical applications for MU-MIMO in WLAN. Using collected channel measurements, the trade-off between CSI accuracy and the effective transmission time has been evaluated. and The optimal sounding interval is analysed for the given channel environment. For practical scenarios where the indoor radio environment can change unpredictably, a low-complexity dynamic sounding approach is proposed that updates the sounding interval to improve the effective throughput. Reference throughput is selected as the metric to determine the sounding interval. Under the collected channel measurements, it is shown that significant throughput improvements (up to 31.8%) for IEEE 802.11ac systems can be achieved by using the proposed approach, especially in the highly dynamic environments. This work can be extended to analyze other commercial MU-MIMO / massive MIMO systems and derive effective channel sounding solutions for new radio bands in heterogeneous radio environments.

## References

[1] J. Kim and I. Lee, "802.11 WLAN: history and new enabling MIMO techniques for next generation standards," *IEEE Commun. Mag.,* vol. 53, pp. 134-140, 2015.

[2] W. Yoo, Y. Jung, M. Y. Kim, and S. Lee, "A pipelined 8-bit soft decision viterbi decoder for IEEE802.11ac WLAN systems," *IEEE Trans. Consumer Electron.,* vol. 58, pp. 1162-1168, 2012.

[3] H. Kasai, "Time-slot based event-driven network switch control for information sharing in multiple WLANs," *IEEE Trans. Consumer Electron.,* vol. 59, pp. 521-529, 2013.

[4] F. Birlik, O. Gurbuz, and O. Ercetin, "IPTVhome networking via 802.11 wireless mesh networks: an implementation experience," *IEEE Trans. Consumer Electron.,* vol. 55, pp. 1192-1199, 2009.

[5] V. Jones and H. Sampath, "Emerging technologies for WLAN," *IEEE Commun. Mag.,* vol. 53, pp. 141-149, 2015.

[6] E. Perahia, "IEEE 802.11n development: history, process, and technology," *IEEE Commun. Mag.,* vol. 46, pp. 48-55, 2008.

[7] R. Nee, "Breaking the gigabit-per-second barrier with 802.11ac," *IEEE Wireless Commun.,* vol. 18, pp. 4-4, 2011.

[8] D. Deng, K. Chen, and R. Cheng, "IEEE 802.11ax: Next generation wireless local area networks," in *10th International Conference on Heterogeneous Networking for Quality, Reliability, Security and Robustness*, Rhodes, Greece, 2014, pp. 77-82.

[9] G. Dimic and N. D. Sidiropoulos, "On downlink beamforming with greedy user selection: performance analysis and a simple new algorithm," *IEEE Trans. Signal Process.,* vol. 53, pp. 3857-3868, 2005.

[10] T. Yoo and A. Goldsmith, "On the optimality of multiantenna broadcast scheduling using zero-forcing beamforming," *IEEE J. Sel. Areas Commun.,* vol. 24, pp. 528-541, 2006.

[11] O. Aboul-Magd, U. Kwon, Y. Kim, and C. Zhu, "Managing downlink multi-user MIMO transmission using group membership," in *IEEE Consumer Communications and Networking Conference*, Las Vegas, USA, 2013, pp. 370-375.

[12] M. X. Gong, E. Perahia, R. Stacey, R. Want, and S. Mao, "A CSMA/CA MAC protocol for multi-user MIMO wireless LANs," in *IEEE Global Telecommunications Conference*, Miami, USA, 2010, pp. 1-6.

[13] V. Valls and D. J. Leith, "Proportional fair MU-MIMO in 802.11 WLANs," *IEEE Wireless Commun. Lett.,* vol. 3, pp. 221-224, 2014.

[14] A. Adhikary, H. C. Papadopoulos, S. A. Ramprashad, and G. Caire, "Multi-user MIMO with outdated CSI: training, feedback and scheduling," in *49th Annual Allerton Conference on Communication, Control, and Computing*, Monticello, USA, 2011, pp. 886-893.

[15] N. Jindal and S. Ramprashad, "Optimizing CSI feedback for MU-MIMO: tradeoffs in channel correlation, user diversity and MU-MIMO efficiency," in *IEEE 73rd Vehicular Technology Conference*, Budapest, Hungary, 2011, pp. 1-5.

[16] H. V. Balan, R. Rogalin, A. Michaloliakos, K. Psounis, and G. Caire, "Achieving high data rates in a distributed MIMO system," in *Proc. of the 18th Annual International Conference on Mobile Computing and Networking*, Istanbul, Turkey, 2012, pp. 41-52.

[17] X. Xie, X. Zhang, and K. Sundaresan, "Adaptive feedback compression for MIMO networks," in *Proc. of the 19th Annual International Conference on Mobile Computing & Networking*, Miami, USA, 2013, pp. 477-488.

[18] E. Perahia and R. Stacey, *Next generation wireless LANs: 802.11 n and 802.11 ac*: Cambridge University Press, 2013.

[19] M. Gast, *802.11 ac: A survival guide*: O'Reilly Media, Inc., 2013.

[20] L. Zhang, L. Song, M. Ma, and B. Jiao, "On the Minimum Differential Feedback for Time-Correlated MIMO Rayleigh Block-Fading Channels," *IEEE Trans. on Commun.,* vol. 60, pp. 411-420, 2012.

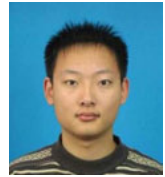

**Xiaofu Ma** (S'13) received the B.S. degree in electronics from Northwest University, Xi'an, China, in 2008, and the M.S. degree in computer science from Tongji University, Shanghai, in 2011, and is currently pursuing the Ph.D. degree in electrical engineering at Virginia Polytechnic Institute and State University, Blacksburg, USA. His research interests include network protocol and optimization, cognitive radio networks, spectrum sharing, chipless RFID, wireless healthcare and wireless local area networks.

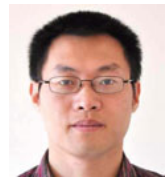

**Qinghai Gao** received his B.S. and M.S. from Xidian University, Xi'an, China, in 1999 and 2002 respectively, and the Ph.D. degree from Arizona State University in 2008. He is currently with Qualcomm Atheros, San Jose, CA. His research interests are in the areas of cross-layer optimization and cooperative communications in wireless networks.

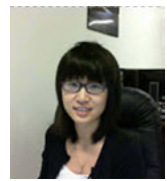

**Ji Wang** received the B.S. degree in Electrical Engineering from Harbin Institute of Technology, Harbin, China, in 2009, and the M.S. degree in applied mathematics and statistics with electrical engineering minor in 2012 from University of Minnesota, and is currently pursuing the Ph.D. degree in electrical engineering at Virginia Polytechnic Institute and State University, Blacksburg, USA. Her research interests include cognitive radio network protocol design and analysis of such systems and networks using game theory.

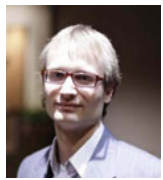

**Vuk Marojevic** received the Dipl.-Ing. degree from the University of Hannover in 2003 and the Ph.D. degree from the Universitat Politècnica de Catalunya (UPC), Spain, in 2009, both in electrical engineering. He is currently with the Wireless @ Virginia Tech research group. His research interests include software-defined and cognitive radio technology, the long-term evolution (LTE), wireless testbeds, infrastructure and waveforms for reliable high-capacity networks.

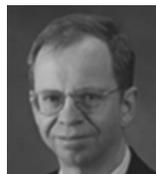

**Jeffrey H. Reed** (F'05) received the B.S.E.E., M.S.E.E., and Ph.D. degrees from the University of California Davis, Davis, CA, USA, in 1979, 1980, and 1987, respectively. He is the founder of Wireless @ Virginia Tech, and served as its director until 2014. He is the Founding Faculty member of the Ted and Karyn Hume Center for National Security and Technology and served as its interim Director when founded in 2010. He is cofounder of Cognitive Radio Technologies (CRT), a company commercializing of the cognitive radio technologies; Allied Communications, a company developing technologies for 5G systems; and for Power Fingerprinting, a company specializing in security for embedded systems. Dr. Reed is a Distinguished Lecturer for the IEEE Vehicular Technology Society. In 2012, Dr. Reed served on the President's Council of Advisors of Science and Technology Working Group that examines ways to transition federal spectrum to allow commercial use and improve economic activity.